\documentclass[twocolumn,showpacs,amsmath,amsfonts,floatfix,aps]{revtex4}
\usepackage{graphicx}
\usepackage{color}

\begin{document}
\title{Spectra of extended systems from Reduced Density Matrix Functional Theory}
\author{S. Sharma$^{1}$}
\email{sharma@mpi-halle.mpg.de}
\author{S. Shallcross$^{2}$}
\author{J. K. Dewhurst$^{1}$} 
\author{E. K. U. Gross$^{1}$}
\affiliation{1 Max-Planck-Institut f\"ur Mikrostrukturphysik, Weinberg 2, 
D-06120 Halle, Germany.}
\affiliation{2 Lehrstuhl f\"ur Theoretische Festk\"orperphysik,
Staudstr. 7-B2, 91058 Erlangen, Germany.}

\date{\today}

\begin{abstract}
We present a method for calculating the spectrum of periodic solids within
reduced density matrix functional theory. An application of this method to the
strongly correlated transition metal oxide series demonstrates
that (i) an insulating state is found in the absence of magnetic order
and, in addition, (ii) the interplay between the charge transfer
and Mott-Hubbard correlation is correctly described. In this respect we find
that while NiO has a strong charge transfer character to the electronic gap,
with substantial hybridization between $t_{2g}$ and oxygen-$p$ states
in the lower Hubbard band, for MnO this is almost entirely absent. 

\end{abstract}

\maketitle

%%%%%%%%%%%%%%
% Introduction
%%%%%%%%%%%%%%

A derivate of the ground-state density functional theory (DFT) calculations 
are the Kohn-Sham (KS) eigenvalues, which lead to a \emph{non-interacting} spectrum.
Even though the KS equations represent an auxiliary non-interacting system  
whose states and eigenvalues may be quite different from the true quasi-particle
system,
empirical evidence shows that in many cases this single particle KS
spectrum is in agreement with the x-ray photo-emission Spectroscopy (XPS) and 
Bremsstrahlung isochromat spectroscopy (BIS) experiments \cite{bis,elp91a,elp91b,sawat84}. 
However, for strongly correlated materials, this KS spectrum is in fundamental 
disagreement with experimental reality. In the absence of spin-ordering 
all modern
exchange correlation (xc) functionals within DFT fail 
to predict an insulating ground-state for
transition metal mono-oxides (TMOs), the prototypical Mott insulators. On the 
other hand, it is well 
known experimentally that these materials are insulating in nature even at 
elevated temperatures (much above the  N\'eel temperature) \cite{tjer,jauch},
indicating that the magnetic order is not the driving mechanism for the existence 
of gap, but instead is a co-occurring phenomenon. 
%In fact not only DFT, but most 
%modern many-body techniques like $GW$ method also fail\cite{rodl09,arya95} to
%capture the insulating behavior in TMOs without explicit long range spin 
%ordering. 

In this regard reduced density matrix functional theory (RDMFT) has 
proved to be valuable in that it not only improves 
upon the KS band gaps for insulators in general, but also predicts TMOs as 
insulators, even in the absence of long range spin-order\cite{sharma08}. This 
clearly points towards its ability to capture the Mott-localization physics. 
Despite this success the effectiveness of RDMFT as ground-state theory
is seriously hampered by the absence of a
technique for the determination of spectral information.
In this work, we present a technique for calculating the spectrum 
within the framework of RDMFT, finding good agreement with experiment
for a selection of TMO's. We father validate this method by a comparison
of the subtle $t2_g$ and $e_g$ irreducible DOS ordering between RDMFT and the well
established $GW$ and Dynamical Mean Field Theory (DMFT) methods.

%%%%%%%%%%%%%%
% RDMFT theory
%%%%%%%%%%%%%%
Within RDMFT, the one-body reduced density 
matrix (1-RDM) is the basic variable \cite{lodwin,gilbert} 
\begin{align}\label{1rdm}
 \gamma({\bf r}, {\bf r'})\equiv N\int d^3r_2\ldots d^3r_N
 \Psi({\bf r},{\bf r}_2 \ldots {\bf r}_N)
 \Psi^*({\bf r}',{\bf r}_2 \ldots {\bf r}_N),
\end{align}
where $\Psi$ denotes the many-body wavefunction and $N$ is the total number of
electrons. Diagonalization of $\gamma$ produces a set of orthonormal Bloch functions, 
the so called natural orbitals\cite{lodwin}, 
$\phi_{i{\bf k}}$, and occupation numbers, $n_{i{\bf k}}$, leading to the
spectral representation
\begin{align}\label{1rdm-spect}
 \gamma({\bf r},{\bf r}')=\sum_{i{\bf k}} 
 n_{i{\bf k}}\phi_{i{\bf k}}({\bf r})\phi_{i{\bf k}}^*({\bf r}'),
\end{align}
where the necessary and sufficient conditions for ensemble  $N$-representability
of $\gamma$ \cite{coleman} require $0\le n_{i{\bf k}}\le 1$ for all $i$ and ${\bf k}$, 
and $\sum_{i{\bf k}}n_{i{\bf k}}=N$.

In terms of $\gamma$, the total ground-state energy \cite{gilbert} of the 
interacting system is (atomic units are used throughout)
\begin{align} \label{etot} \nonumber
E[\gamma]=&-\frac{1}{2} \int\lim_{{\bf r}\rightarrow{\bf r}'}
\nabla_{\bf r}^2 \gamma({\bf r},{\bf r}')\,d^3r'
+\int\rho({\bf r}) V_{\rm ext}({\bf r})\,d^3r \\
&+\frac{1}{2}  \int 
\frac{\rho({\bf r})\rho({\bf r}')}
{|{\bf r}-{\bf r}'|}\,d^3r\,d^3r'+E_{\rm xc}[\gamma],
\end{align}
where $\rho({\bf r})=\gamma({\bf r},{\bf r})$, $V_{\rm ext}$ is a given
external potential, and $E_{\rm xc}$ we call the xc dos
energy functional. In principle, Gilbert's \cite{gilbert} generalization of the
Hohenberg-Kohn theorem to the 1-RDM guarantees the existence of a functional
$E[\gamma]$  whose minimum, for fixed $V_{\rm ext}$ yields the exact $\gamma$ 
and the exact ground-state energy of systems characterized by the external 
potential $V_{\rm ext}({\bf r})$. In practice, however, the correlation energy 
is an unknown functional of $\gamma$ and needs to be approximated. While there 
are several known approximations
for the xc energy functional, the most promising for extended systems is the
power functional\cite{sharma08} where the xc energy reads
\begin{align} \label{power} 
&E_{\rm xc}[\gamma]= -\frac{1}{2}\int \, \int d^3r' d^3r
 \frac{|\gamma^{\alpha}({\bf r},{\bf r}')|^2}{|{\bf r}-{\bf r}'|}
\end{align}
%%%%%%%%%%%%%%%%%
% choice of alpha
%%%%%%%%%%%%%%%%%
where $\alpha$ is a system dependent parameter\cite{power_finite,sharma08,esa11}. 
However, in the present work we fix the value of $\alpha=0.656$ for all TMOs.

%%%%%%%%%%%%%%%%%%%%%%%%%%%%%%%%%%%%
% connection 2 spectral function  %%
%%%%%%%%%%%%%%%%%%%%%%%%%%%%%%%%%%%%
In the following we first devise a theoretical method to obtain an expression for the spectral density function with RDMFT, 
which by its very nature, is a ground-state theory and then further apply this method to the case of TMOs. We start from the 
definition of the Green's function written in the basis of the natural orbitals,
\begin{align}\label{GF}
iG_{\alpha \beta}(t-t')=\frac{1}{\langle \Psi_0^N |\Psi_0^{N} \rangle }
\langle \Psi_0^N | T[a_{\alpha}(t) a^{\dagger}_{\beta}(t')] | \Psi_0^{N} \rangle,
\end{align}
where $\alpha \equiv \{i,{\bf k}\}$ with the index $i$ labeling the orbital for a given {\bf k} and
$a$,$a^{\dagger}$ are the creation and annihilation operators
associated with the complete set of natural orbitals. Inserting in Eq. (\ref{GF}) the completeness relation for a
restricted but physically significant\cite{lodwin} set of ($N\pm1$)-particle states,
\begin{align}\nonumber
|\Psi_{\zeta}^{N+1} \rangle=\frac{1}{\sqrt{n_{\zeta}}}a^{\dagger}_{\zeta} | \Psi_0^N \rangle, \, \, \, \, \, \,
|\Psi_{\zeta}^{N-1} \rangle=\frac{1}{\sqrt{(1-n_{\nu})}}a_{\zeta} | \Psi_0^N \rangle,
\end{align}\label{Np1st}
the imaginary part of the Green's function (spectral density function) can be expressed as:
\begin{align}\nonumber
&A_{\alpha\beta}(\omega)= 2\pi\sum_{\zeta}\frac{1}{n_{\zeta}}\langle \Psi_0^N | a_{\alpha} a^{\dagger}_{\zeta}| \Psi_0^{N} \rangle 
\langle \Psi_0^{N} | a_{\zeta} a^{\dagger}_{\beta} | \Psi_0^{N} \rangle \delta(\omega-\epsilon^+_{\zeta}) \\ 
&-2\pi\sum_{\nu}\frac{1}{1-n_{\nu}}\langle \Psi_0^N | a^{\dagger}_{\alpha} a_{\nu}| \Psi_0^{N} \rangle 
\langle \Psi_0^{N} | a^{\dagger}_{\nu} a_{\beta} | \Psi_0^{N} \rangle 
 \delta(\omega-\epsilon^-_{\nu})
\end{align}\label{spec-dens}
with $\epsilon^{\pm}_{\nu}=E_0^N - E_{\nu}^{N\pm1}$. The trace of this quantity is usually called the density of 
states (DOS), and in this basis of natural orbitals this assumes a simple form:
\begin{align}\label{dos}
{\rm DOS} = \sum_{\zeta} n_{\zeta} \delta(\omega-\epsilon^+_{\zeta})+ 
\sum_{\nu} (1-n_{\nu}) \delta(\omega-\epsilon^-_{\nu}),
\end{align}
where the first term gives the occupied part of the spectrum and second the unoccupied part.

Now what remains is to calculate the excitation energies $\epsilon^{\pm}_{\nu}=\epsilon^{\pm}_{i \bf k}=E_0^N - E_{i\bf k}^{N\pm1}$, where 
$E_{i \bf k}(N \pm 1)$ is the energy of the system with an electron, with specific momentum {\bf k}, added/removed; 
this energies are accessible within RDMFT because systems with an 
added/removed particle can be viewed as the ground-state energy of 
a $(N \pm 1)$-electron system constrained to have total momentum {\bf k}. 
While in experiments $E_{\bf k}(N \pm 1)$ represents 
the total energy of a macroscopic block of material, in the theoretical description $E_{\bf k}(N \pm 1)$ is total
energy of a large but periodically repeated Born-von Karman (BvK) cell, where a constant charge background is added
to keep the total (infinite) system charge neutral. 
Since the total energies for BvK cell are computationally  very demanding to calculate, we introduce a simplification 
which is \emph{not conceptual} in nature but rather a numerical trick similar to the Slater transition state procedure \cite{slater}: 
we first introduce total ground-state
energies, $E_{\bf k}(N \pm \eta)$, where a fractional number of particles, $\eta$, has been added/subtracted at a given {\bf k}.
These energies can be defined as proper ensemble energies of $N$ and $N \pm 1$ particle systems \cite{perdew}. Then following Slater, 
the total energy difference, $\epsilon^{\pm}_{\nu}$,  can be approximated as 
\begin{align}\label{dedeta}
 \epsilon^{\pm}({\bf k})= 
 \left.\frac{\partial E_{\bf k}(N \pm \eta)}{\partial \eta} \right|_{\eta=1/2},
\end{align}

\begin{figure}[ht]
\centerline{\includegraphics[width=\columnwidth,angle=-0]{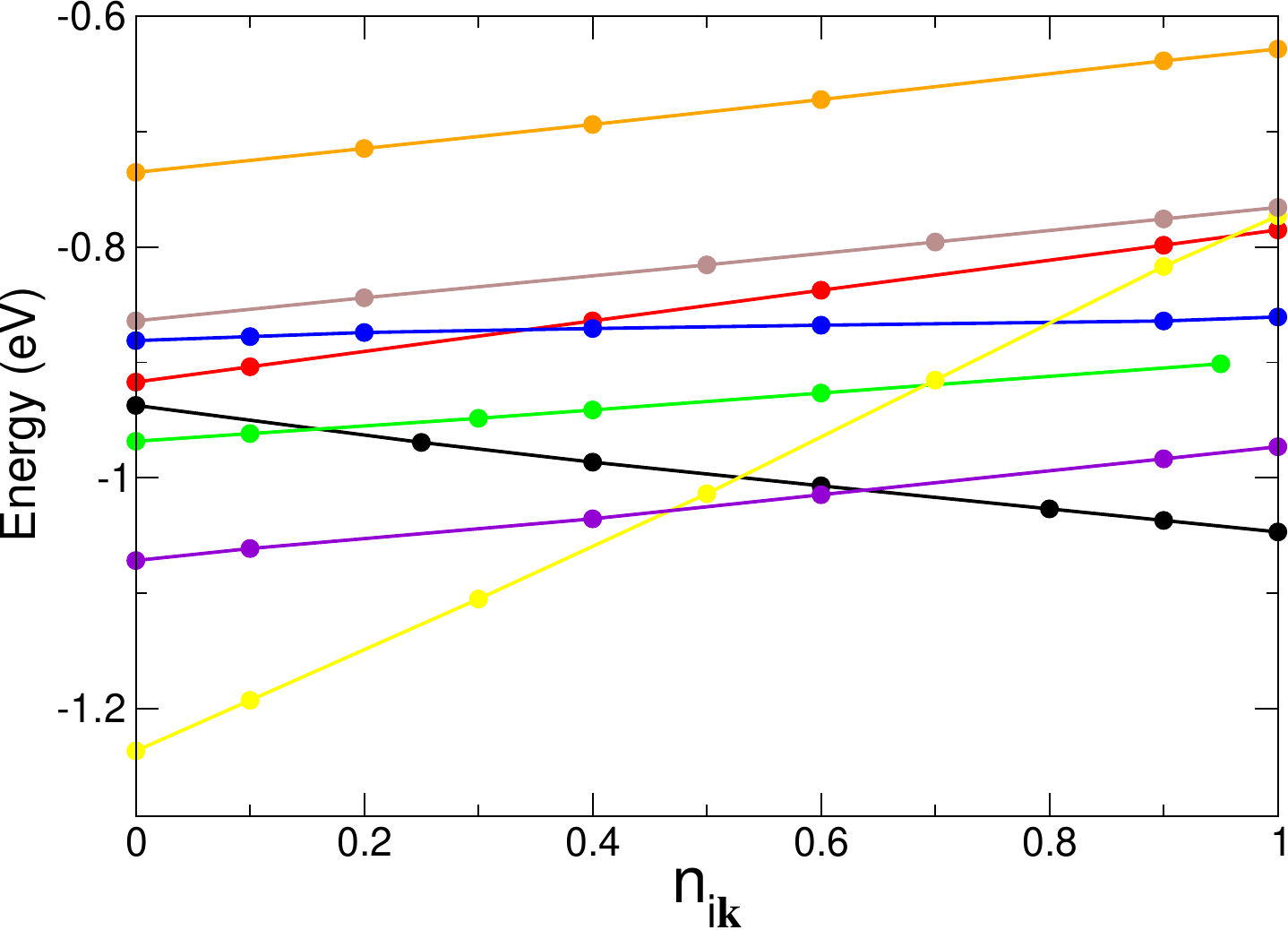}}
\caption{(Color online) Change in total energy upon changing a single occupation
number $n_{i{\bf k}}$. Results are calculated for various $i{\bf k}$
for NiO (black and red), CoO (green and blue), MnO (yellow and brown) and FeO(orange and violet).}
\vspace{0.7cm}
\label{stline}
\end{figure}
In order to calculate $\epsilon^{\pm}$ as expressed in Eq. (\ref{dedeta}) one requires number of {\bf k}-points times
the number of natural orbital (typically $\sim$2500) ground-state calculations. This is still a formidable task and 
hence we make another simplification; we assume that upon adding/subtracting an electron at {\bf k} from the BvK cell
the only occupation number that will change significantly is the one that corresponds to the very same {\bf k}
while all the other occupation numbers and natural orbitals remain unchanged. 
%Clearly this will not be true for finite systems, but for infinite solids the assumption is justified. 
Under this assumption Eq. \ref{dedeta} reduces to
\begin{align}\label{dedn}
 \epsilon^{\pm}({\bf k})=\left.\frac{\partial E[\{\phi\},\{n\}]}
 {\partial n_{\bf k}}\right|_{n_{{\bf k}=1/2}}
\end{align}
This approximation can be further validated by plotting E as a function of $n_{\bf k}$--
we find for all the materials involved a nearly linear behaviour (see Fig. \ref{stline}). This implies that the 
Slater-type evaluation of the total-energy difference in Eq. (\ref{dedn}) is rather accurate.
While for the highest occupied and the lowest unoccupied QP state the above procedure is perfectly justified, we use it also
for higher/lower lying states, i.e. we calculate the spectrum using Eq. (\ref{dos}) with
\begin{align}\label{dedn2}
 \epsilon^{\pm}(i,{\bf k})=\left.\frac{\partial E[\{\phi\},\{n\}]}
 {\partial n_{i{\bf k}}}\right|_{n_{i{\bf k}=1/2}}
\end{align}
Use of the ground-state RDMFT in Eq. (\ref{dedn2}) for states away from the chemical potential can be problematic; 
the procedure implicitly assumes that local-minima of the ground-state functional represent excited-state energies, 
a feature that has been shown for the ground-state DFT functional\cite{perdew2}. Whether a similar statement can 
be proved in RDMFT is currently unknown. 

%%%%%%%%%%%%%%%%%%%
% Technical details
%%%%%%%%%%%%%%%%%%%
Following the above procedure the DOS for the strongly correlated Mott insulators
NiO, CoO, FeO and MnO is calculated using the full-potential linearized augmented plane wave code Elk\cite{elk},
with practical details of the calculations following the scheme described in Ref.~(\onlinecite{sharma08}).

%%%%%%%%%%%%%%%%%%%%%%%%%%%%%
% Results and discussion: TMO
%%%%%%%%%%%%%%%%%%%%%%%%%%%%%
\begin{figure}[ht]
\vspace{0.5cm}
\centerline{\includegraphics[width=\columnwidth,angle=-0]{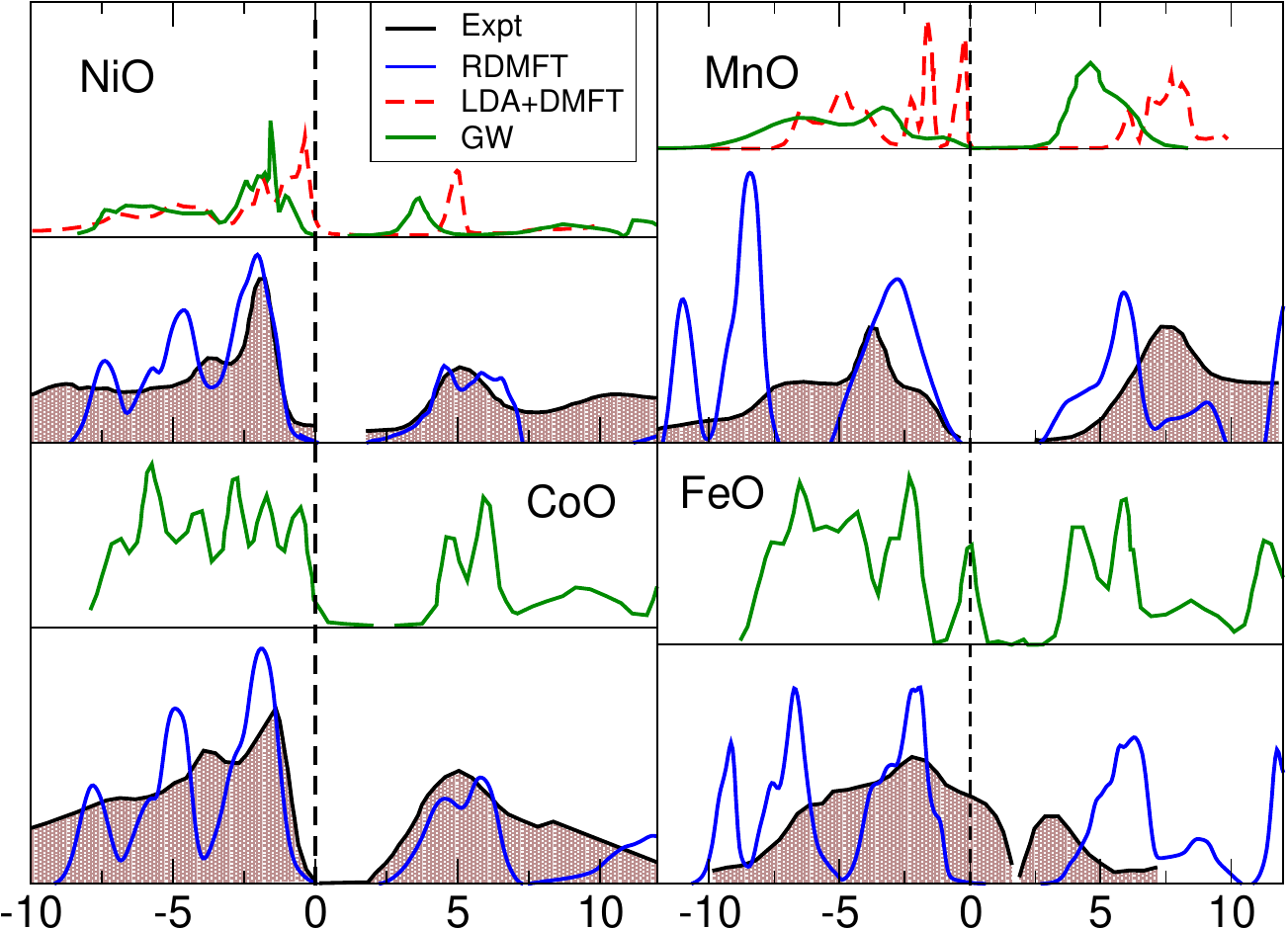}}
\caption{(Color online) Density of states for the TMOs.
Shown are XPS and BIS spectra, in addition
to calculations using the $GW$, DMFT, and RDMFT methods.
The $GW$ and DMFT results are from spin-polarized calculations, and 
are vertically shifted for clarity, while the RDMFT calculations are spin-unpolarized with $\alpha=0.656$ for
all materials.}
\vspace{0.7cm}
\label{tmo-dos}
\end{figure}

Presented in Fig.~\ref{tmo-dos} are the spectra generated
via Eq.~(\ref{dedn2}) for the Mott
insulators under consideration. Also shown are $GW$ data taken from
Refs.~\onlinecite{rodl09,kobayashi08} and DMFT results form
Refs.~\onlinecite{oki08,ren06,kunes08}. For details of these
calculations we refer the reader to the aforementioned
works, however we note that both DMFT and $GW$ method require 
as a starting point the spin-polarized DFT. Additionally DMFT
also requires an empirical Mott-Hubbard parameter $U$ \cite{anisomov}. 
The experimental data shown in Fig. ~\ref{tmo-dos} are taken from 
Refs.~\onlinecite{elp91a,elp91b,sawat84,bagus77,bowen75}.

It is immediately apparent from Fig.~\ref{tmo-dos} that
RDMFT captures the essence of Mott-Hubbard physics: all
the TMOs considered are \emph{insulating in the absence of any long range spin
order}. This fact was already noticed in the previous work \cite{sharma08} 
where the presence of gap without any spin-order was deduced via a very 
different technique, namely the discontinuity in the chemical potential as a 
function of the particle number.

A closer examination of the spectra for NiO and
CoO reveals an excellent agreement between the RDMFT peaks
and the corresponding XPS and BIS data. In fact, not only
the peak positions, but also their relative weights are well reproduced. 
For MnO one notes that the agreement between
experiment and RDMFT, regarding the relative weights of the peaks, is somewhat worse.
Turning to the case of FeO, it must be recalled that Fe segregation,
unavoidable in this compound, precludes the experimental realization
of pure FeO samples. For this reason the only existing experimental
data are rather old, and the presumably substantially contaminated and broadened
data present no distinct features that may be used for comparison.

One notes that the agreement between experiments and RDMFT DOS is best for
NiO which has the lowest magnetic moment (1.9 $\mu_B$) amongst
the TMOs considered here, and the worst for MnO which has the largest
moment (4.7 $\mu_B$). As the RDMFT calculations presented here are
non-magnetic (i.e., spin degenerate) the trend is natural, and indicates that
for the large moment TMOs the co-occurring magnetic order does contribute
significantly to the spectral density, a fact we will demonstrate later 
by performing spin-polarized calculations.

Turning to a comparison of the RDMFT spectra with the corresponding $GW$ and DMFT results, one 
notes that for NiO all three methods are in close agreement. For
MnO the $GW$ method incorrectly leads to semi-metallic behaviour, but both RDMFT and DMFT, 
as in experiment, show an insulating character. The actual values of the insulating gaps that 
may be extracted from Fig.~\ref{tmo-dos} are 2.3 eV (4.3 eV), 2.3 eV (2.8 eV), 2.9 eV (2.4 eV),
and 2.5 eV (3.6 eV), for NiO, CoO, FeO, and MnO respectively with the corresponding experimental gap 
given in parenthesis.

%%%%%%%%%%%%%%%%%%%%%%%%%%%
% Results for magnetic DOS
%%%%%%%%%%%%%%%%%%%%%%%%%%%
\begin{figure}[ht]
\centerline{\includegraphics[width=\columnwidth,angle=-0]{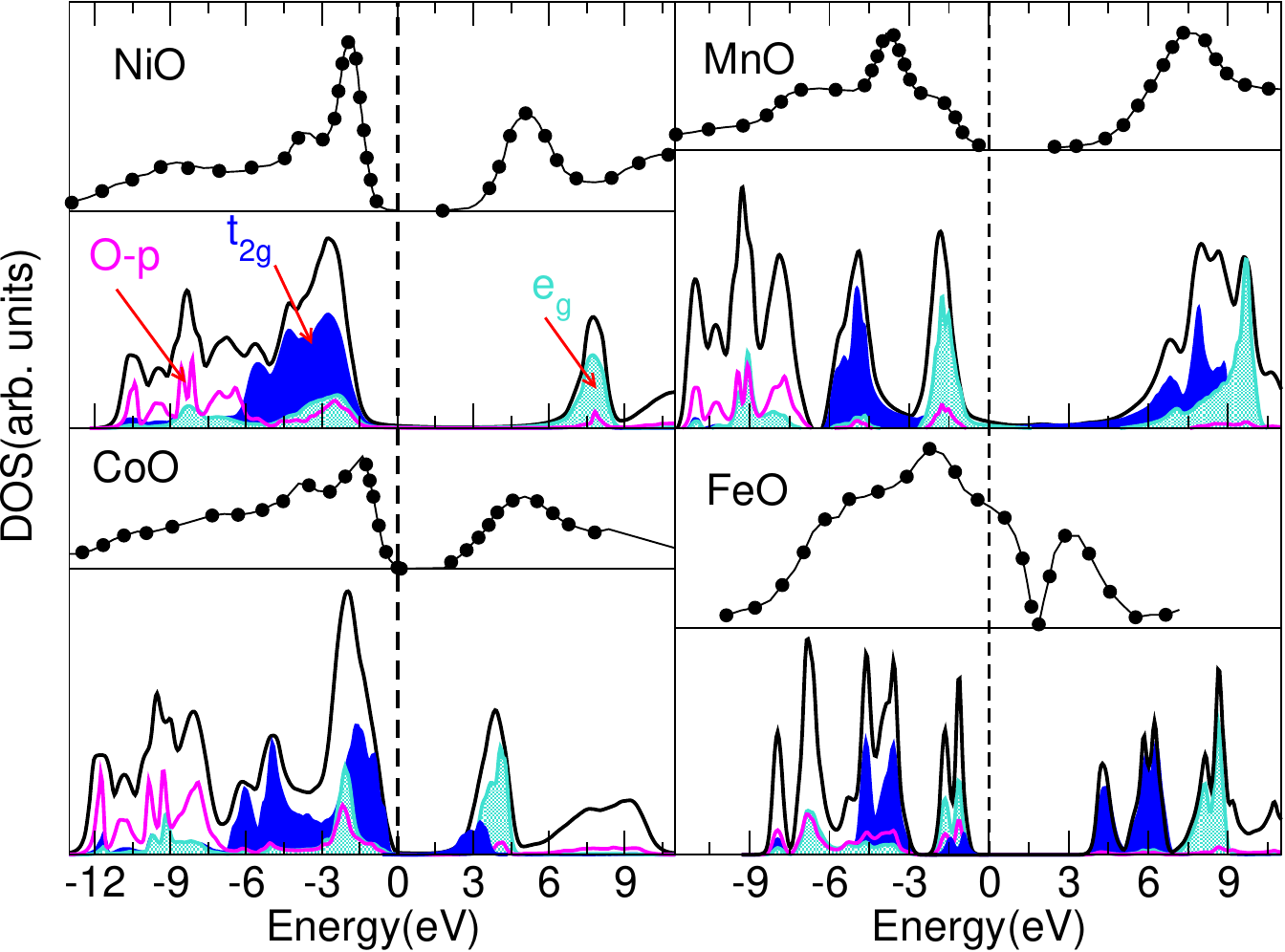}}
\caption{(Color online) Density of states for the TMOs in presence of AFM order.
Site and angular momentum projected DOS are also presented for transition metal $e_g$ and $t_{2g}$
states and Oxygen-$p$ states. In addition XPS and BIS spectra (shifted up for clarity) are
presented for comparison. Again, $\alpha=0.656$ for all materials. }
\label{tmo-mag}
\end{figure}
It might be argued that the zero temperature ground-state for all these TMOs
has long range anti-ferromagnetic (AFM) order and RDMFT might not reproduce similar good
results for DOS when such an ordering is invoked. To clarify this we extend the power-functional
in Eq. (\ref{power}) to the magnetic case by treating the natural orbitals as Pauli-spinors. 
The results thus obtained are shown in Fig. (\ref{tmo-mag}). 

Reassuringly, we find that for NiO and CoO the inclusion of long range AFM spin order
only brings the already good results into closer quantitative agreement with experiment: the band
gaps increase to 4.5eV (4.3eV) and 2.6eV (2.8eV) respectively, with the experimental gaps in parenthesis.
For MnO and FeO, however, the changes upon invoking spin order are dramatic. The DOS  
changes significantly in both cases, with the detailed comparison of peak structure now in good agreement with experiments.
The value of the local moments we find to be 1.36(1.9)$\mu_B$, 2.7(3.3)$\mu_B$, 3.35(3.32)$\mu_B$ 
and 3.38(4.7)$\mu_B$ for NiO, CoO, FeO and MnO respectively, again with the experimental values in parenthesis.

%%%%%%%%%%%%%%%%%%%%
% Results for PDOS
%%%%%%%%%%%%%%%%%%%%
As is well known, while the insulating state of TMOs is driven by 
a charge localization due to strong Coulomb repulsion (Mott-Hubbard correlation),
an important auxiliary mechanism is charge transfer\cite{zaanen85}
due to hybridization between ligand and transition metal (TM) states.
Amongst the TMO series this latter mechanism is generally believed 
to play an important role in the case of NiO, but to be of decreasing importance
as the atomic number is lowered, with the insulating state of MnO thought to be
driven entirely by Mott-Hubbard correlation. Clearly, an outstanding challenge for any 
\emph{ab-initio} theory is to capture \emph{both} these aspects
of TMO physics. 

In Fig.~\ref{tmo-mag} we also present the site and angular momentum projected
DOS for the TMOs considered in this work. The electronic gap, as expected,
always occurs between lower and upper Hubbard bands dominated by transition metal d-states. 
However, while for NiO one finds a significant component of oxygen-$p$ states in the 
lower Hubbard band, for the other TMOs this hybridization between oxygen-$p$ 
and TM-$d$ states reduces, and is almost absent in the case of
MnO, indicating that for this material the insulating state is driven
mostly by Mott-Hubbard correlations. 

As a validation of our method for calculation of the DOS we may compare these features of the projected DOS, and
in particular the ordering in energy of the $t_{2g}$ and $e_g$ states, with well established 
\emph{ab-initio} many-body techniques
such as DMFT and the $GW$ method\cite{kunes08,ren06,rodl09}. In all cases we find an excellent agreement, signaling
that the method we present here yields not merely gross spectral features, but an accurate description of
detailed and subtle features of the resolved state density.

%%%%%%%%%%%%%%%%%%%%%%%%%%%%%%%%%%%%%%%%%
% Results and discussion: Charge density
%%%%%%%%%%%%%%%%%%%%%%%%%%%%%%%%%%%%%%%%%
\begin{figure}[ht]
\centerline{\includegraphics[width=\columnwidth,angle=-0]{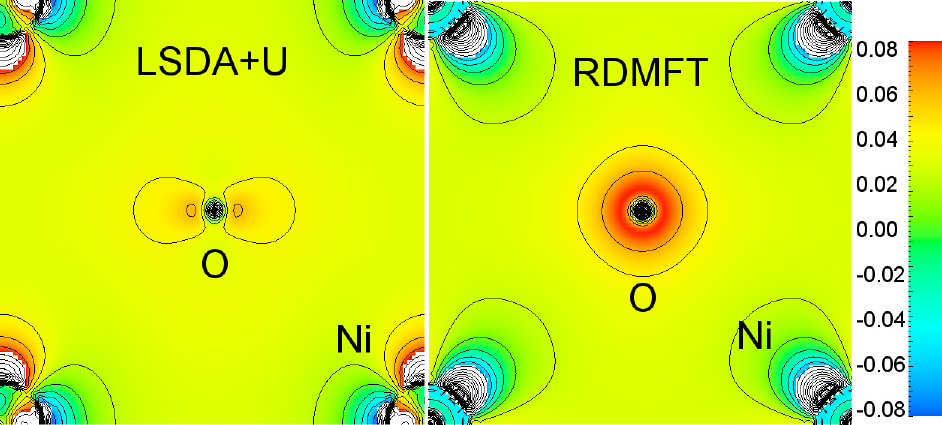}}
\caption{(Color online) Difference between the LSDA charge density and the 
charge densities calculated using LSDA+$U$ and RDMFT, 
($\rho({\bf r})-\rho_{LSDA}({\bf r})$) for 
NiO. Positive values indicate localization of charge as compared to LSDA.}
\label{rho2d}
\end{figure}
A change in the nature of bonding as well as localization of charge as a result of better treating
correlations may be seen in the charge density difference $\rho({\bf r})-\rho_{LSDA}({\bf r})$, 
shown in Fig.~\ref{rho2d} for RDMFT and LSDA+$U$ calculations of NiO. A comparison with LSDA+$U$ is 
instructive as this method (with an appropriate choice of $U$) is able to 
accurately reproduce the insulating gaps of the TMO series and via $U$ adds correlations beyond LSDA. 
Interestingly, one observes an almost spherical charge accumulation at the oxygen 
site, a result in agreement with experiment\cite{dudarev00}, but different
from that found in the corresponding LSDA+$U$ result.

%%%%%%%%%%%%%
% Conclusions
%%%%%%%%%%%%%
To conclude we have presented a method to calculate photo electron
spectra within the framework of RDMFT based on the derivative of the total 
energy with respect to occupation number at half filling. We have shown that the spectral 
information obtained in this way gives a detailed account of the strongly correlated
nature of the TMOs, including the subtle interplay between Mott-Hubbard
correlation and charge-transfer character in these materials. We validate this method by
not only by the agreement with experiment for gross spectral features, but also by 
a detailed comparison of the angular momentum resolved partial DOS for TMO series
with that of well established many-body techniques, in all cases finding excellent agreement.

%\bibliography{spectra}

\end{document}